# Multipath Error Correction in Radio Interferometric Positioning Systems

Cheng Zhang, Wangdong Qi*, *Member, IEEE*, Li Wei, Jiang Chang, and Yuexin Zhao

*Abstract*—The radio interferometric positioning system (RIPS) is an accurate node localization method featuring a novel phase-based ranging process. Multipath is the limiting error source for RIPS in ground-deployed scenarios or indoor applications. There are four distinct channels involved in the ranging process for RIPS. Multipath reflections affect both the phase and amplitude of the ranging signal for each channel. By exploiting untapped amplitude information, we put forward a scheme to estimate each channel's multipath profile, which is then subsequently used to correct corresponding errors in phase measurements. Simulations show that such a scheme is very effective in reducing multipath phase errors, which are essentially brought down to the level of receiver noise under moderate multipath conditions. It is further demonstrated that ranging errors in RIPS are also greatly reduced via the proposed scheme.

*Index Terms*—Radio interferometric positioning system, multipath mitigation, phase measurement, amplitude information

## I. INTRODUCTION

THE radio interferometric positioning system (RIPS) is an accurate node localization method featuring a novel ranging process based on phase measurement [1]. A prototype implementation of RIPS on the Mica2 platform [2] achieves an average localization accuracy of 4 cm and a range of 160 m [3]. As a promising approach to low-cost and accurate localization, RIPS has been extended within many variations [4], [5], [6], [7], [8], [9], [10], [11], [12], [13], [14].

The performance of RIPS, however, can suffer severely due to multipath propagation in some environments [15]. For ground-deployed sensor nodes, signals reflected from the ground can cause significant error in phase measurements [3]. Multipath propagation has also been identified as the limiting factor in indoor applications [16].

Some system design choices can help to reduce multipath effects on RIPS. Among those are iterative refinements, elevated antennas, lower carrier frequencies, and redundant infrastructure nodes [3], [16]. However, protracted time frames, bulkier antennas, and higher deployment costs associated with these design choices can be unacceptable within the wireless sensor network or mobile application domains.

To address inherent challenges resulting from multipath

This work was supported by the National Natural Science Foundation of China (Grants 61273047 and 61573376), and the Natural Science Foundation of Jiangsu Province, China (BK20130068).

The authors are with the PLA University of Science and Technology, Nanjing, Jiangsu 210007, China (e-mails: zhangchengnj@gmail.com; wangdongqi@gmail.com; wlnb@hotmail.com; changj10@vip.sina.com; zhaoyxsd@gmail.com).
*Wangdong Qi is the corresponding author.

phenomena, several RIPS variations implement different ranging signals from that of the original scheme. For example, the ranging signal of the dual-tone RIPS uses two tones, as opposed to the singular one used for the original RIPS scheme [13], [14]. The two tones are limited within channel coherence bandwidth; hence, they will ultimately experience comparable channel fading effects. As a result, the differential phase of the two-tone system is immune to multipath. However, dual-tone RIPS no longer supports multiple carrier frequencies from broad bandwidths, which was a crucial feature of RIPS having been responsible for higher accuracy levels [17].

In this letter, we propose a new approach to multipath mitigation in RIPS. By utilizing the untapped amplitude information in ranging signals, we were able to estimate the multipath profile of each involved channel. These multipath profiles were then used to correct corresponding errors in phase measurements.

Simulations show that phase errors are reduced to the level of receiver noise under moderate multipath conditions. They are also significantly reduced under severe multipath conditions. It is also demonstrated that ranging errors due to multipath are greatly reduced via the proposed phase error correction method.

Section II below introduces the system model of RIPS phase measurement in multipath environments. Section III discusses the proposed scheme for multipath error correction. In Section IV, the performance of the proposed scheme is evaluated by simulations, and in Section V, the conclusions are presented.

## II. SYSTEM MODEL

In this section, we first review the theory of RIPS in a benign environment, followed by an examination of the error items induced by multipath during the phase measurement process. Receiver noise is irrelevant to this derivation, and thus, is intentionally omitted in this section.

### A. RIPS in Multipath-Free Environments

The localization of a node in RIPS consists of three stages: phase measurement, range estimation, and location finding.

The basic unit of a phase measurement process in RIPS consists of four nodes, of which two (*A* and *B*) send sinusoids at two close frequencies, $f_A$ and $f_B$ (assuming that $f_A > f_B$). Two other nodes, *C* and *D*, simultaneously measure (i.e., estimate) the phase offset of the two received sinusoids from line-of-sight (LOS) paths. The difference of the two phase offsets $\overline{\varphi_C}$ and $\overline{\varphi_D}$, $\overline{\Delta\varphi} = \overline{\varphi_C} - \overline{\varphi_D}$, is shown to be a constant related to the so-called *q-range* $d_q = d_{AD} - d_{BD} + d_{BC} - d_{AC}$ ($d_{XY}$ is the distance between nodes *X* and *Y*) as [1]

$$\overline{\Delta\varphi} \approx \frac{2\pi f}{c} d_q \pmod{2\pi} \tag{1}$$

where $f = (f_A + f_B)/2$ and $c$ is the speed of radio propagation.

Given the observation of $\overline{\Delta\varphi}$, the q-range can be obtained according to equation (1), albeit with integer ambiguity due to phase wrapping. To resolve any potential ambiguity of the q-range, we measure $\overline{\Delta\varphi}$ at a sequence of measurement frequencies $f_X(k) = f_X(0) + k\Delta f$, $k = 0,1,\ldots,K-1$. From $\overline{\Delta\varphi}(k)$, an optimization procedure can then be used to evaluate the q-range.

Sufficient numbers of q-ranges associated with a node can be used to determine the location of that node in a number of ways. Due to space limitation, we elaborate on neither the determination of the q-range nor the localization of a node in RIPS. We thus refer interested readers to [1], [3], [5] for further information.

Our topic of interest here is potential error in RIPS phase measurements under multipath conditions. For brevity, the index $k$ for measurement frequencies is omitted.

### B. Phase Error under Multipath Conditions

Assume that sender $X$ ($A$ or $B$) transmits a single tone

$$s_X(t) = A_X \exp\left(j(2\pi f_X(t - t_X))\right) \quad (2)$$

where $A_X$ is the amplitude, $f_X$ is the carrier frequency, and $t_X$ is the unknown time instant when $X$ starts to transmit.

Assume that $L_{XY}$ is the number of reflected paths in a channel between nodes $X$ and $Y$, $\alpha_{XY,i}$ is the multipath-to-direct ratio (MDR) of amplitudes of the $i$th reflected signal, $\tau_{XY,i}$ is the difference in delay between the $i$th reflected signal and the direct one, and $\theta_{XY,i}$ is the phase shift caused by reflection in the $i$th reflected path. The receiver $Y$ ($C$ or $D$) then receives the following signal from $X$:

$$r_{XY}(t) = \overline{A_{XY}} \exp\left(j(2\pi f_X t + \overline{\varphi_{XY}})\right)$$
$$\cdot \left(1 + \sum_{i=1}^{L_{XY}} \alpha_{XY,i} \exp\left(-j(2\pi\tau_{XY,i}f_X + \theta_{XY,i})\right)\right) \quad (3)$$

where $\overline{A_{XY}}$ and $\overline{\varphi_{XY}} = -2\pi f_X(t_X + d_{XY}/c)$ are the amplitude and phase of the LOS signal from $X$ to $Y$, respectively. As a composite of multiple single tones with the same frequency, $r_{XY}(t)$ is obviously also a single tone.

Let $r_{XY}(t) = A_{XY} \exp\left(j(2\pi f_X t + \varphi_{XY})\right)$, then

$$A_{XY} = \sqrt{\left(\overline{A_{XY}} + \sum_{i=1}^{L_{XY}} \overline{A_{XY}}\alpha_{XY,i}\cos(2\pi\tau_{XY,i}f_X + \theta_{XY,i})\right)^2 + \left(\sum_{i=1}^{L_{XY}} \overline{A_{XY}}\alpha_{XY,i}\sin(2\pi\tau_{XY,i}f_X + \theta_{XY,i})\right)^2} \quad (4)$$

The phase of the multipath distorted signal is

$$\varphi_{XY} = \overline{\varphi_{XY}} + \epsilon_{XY} \quad (5)$$

Here, the error item $\epsilon_{XY}$ in $\varphi_{XY}$ is induced by reflected paths and is given by

$$\epsilon_{XY} = -\tan^{-1}\frac{\sum_{i=1}^{L_{XY}} \alpha_{XY,i}\sin(2\pi\tau_{XY,i}f_X + \theta_{XY,i})}{1 + \sum_{i=1}^{L_{XY}} \alpha_{XY,i}\cos(2\pi\tau_{XY,i}f_X + \theta_{XY,i})} \quad (6)$$

In the original RIPS, the multipath error term, $\epsilon_{XY}$, is neglected. That is the reason why RIPS does not work satisfactorily in multipath environments.

Since

$$\overline{\Delta\varphi} = \overline{\varphi_{AC}} - \overline{\varphi_{BC}} - \overline{\varphi_{AD}} + \overline{\varphi_{BD}}$$
$$= \varphi_{AC} - \varphi_{BC} - \varphi_{AD} + \varphi_{BD} - (\epsilon_{AC} - \epsilon_{AD} - \epsilon_{BC} + \epsilon_{BD}), \quad (7)$$

a correction to the phase errors caused by all the reflected signals (of the four channels) is necessitated.

### III. PHASE ERROR CORRECTION

In this section, we first look at the overall process, ranging from the received radio signal to the corrected phase measurement. We then discuss, in detail, the multipath parameter estimation process, which is at the heart of the process above. At the end of this section, the effect of phase error correction is illustrated with an example.

#### A. Overview of Phase Error Correction

Our scheme for phase measurement with multipath error correction is shown in Fig. 1.

In down conversion, the radio frequency signal $r_Y(t) = r_{AY}(t) + r_{BY}(t)$ received by node $Y$ is converted to baseband

$$b_Y(t) = r_Y(t) \cdot \exp(-j(2\pi f_Y t + \beta_Y))$$
$$= A_{AY} \exp(j(2\pi(f_A - f_Y)t + \varphi_{AY} - \beta_Y))$$
$$+ A_{BY} \exp(j(2\pi(f_B - f_Y)t + \varphi_{BY} - \beta_Y)) \quad (8)$$

Here, $f_Y$ is the frequency of the local carrier and $\beta_Y$ is the unknown phase shift induced by down conversion.

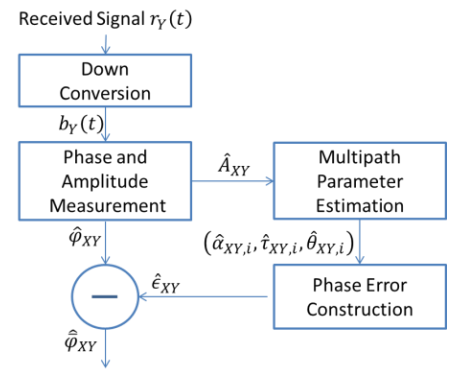

Fig. 1. Phase measurement with multipath error correction

The amplitude and phase of each frequency component of the baseband signal $b_Y(t)$ can be estimated in standard ways. From equation (8), the phase offset of the two tones received by $Y$ is

$$\varphi_Y(t) = (2\pi(f_A - f_Y)t + \varphi_{AY} - \beta_Y)$$
$$- (2\pi(f_B - f_Y)t + \varphi_{BY} - \beta_Y)$$
$$= 2\pi(f_A - f_B)t + \varphi_{AY} - \varphi_{BY} \quad (9)$$

The difference $\Delta\varphi(t)$ between the phase offsets in the baseband signals $b_C(t)$ and $b_D(t)$ is then

$$\Delta\varphi(t) = \varphi_C(t) - \varphi_D(t)$$
$$= \varphi_{AC} - \varphi_{BC} - \varphi_{AD} + \varphi_{BD}$$
$$= \overline{\Delta\varphi} - (\epsilon_{AC} - \epsilon_{AD} - \epsilon_{BC} + \epsilon_{BD})$$
(10)

Therefore, we can extract $\overline{\Delta\varphi}$ from $\Delta\varphi(t)$ once the multipath error items are determined. To determine each multipath error according to equation (6), we need to obtain each channel's multipath profile, which is characterized by the three parameters $\alpha_{XY,i}$, $\tau_{XY,i}$, and $\theta_{XY,i}$.

### B. Multipath Parameter Estimation

A key observation is that amplitude information in the received signal can be utilized to obtain multipath parameters of the channel from transmitter $X$ to receiver $Y$. The subscripts $X$ and $Y$ are omitted in this subsection for brevity.

According to the free-space model, the amplitude of an LOS signal $\bar{A}$ can be expressed as

$$\bar{A} = \frac{c\sqrt{PG}}{4\pi f d}$$
(11)

where $P$ is the transmitted power and $G$ is the antenna gain.

For small $\alpha_i$, the amplitude in equation (4) can be approximated as

$$A = \bar{A} + \sum_{i=1}^{L} \bar{A}\alpha_i \cos(2\pi\tau_i f + \theta_i)$$
(12)

Assuming that multipath parameters, transmitted power, and antenna gain do not change significantly with measurement frequency, we have the $k$th frequency weighted amplitude as

$$A(k)\frac{f(k)}{f(0)} = \bar{A}(0) + \sum_{i=1}^{L} \bar{A}(0)\alpha_i \cos(2\pi\tau_i f(k) + \theta_i)$$
(13)

where $f(k), k = 0,1,\ldots,K-1$ is the sequence of measurement frequencies of node $X$. Note that the right side of equation (13) is periodic with $f$ as the independent variable since it is the superposition of a set of sinusoids and a constant. Therefore, the problem of multipath parameter estimation is essentially equivalent to that of multiple tone parameter estimation, in which the parameter $\tau_i$ acts as "the frequency of the $i$th tone". A simple solution to the estimation problem is given as follows:

*1) Estimation of $\bar{A}(0)$*

An estimator of $\bar{A}(0)$ is given in [18]

$$\hat{\bar{A}}(0) \approx \frac{1}{K}\sum_{k=1}^{K-1} A(k)\frac{f(k)}{f(0)}$$
(14)

*2) Estimation of $\tau_i$*

Defining
$$A_{MP}(k) = A(k)\frac{f(k)}{f(0)} - \hat{\bar{A}}(0),$$
(15)

we can estimate $\tau_i$ with a frequency estimation method based on discrete Fourier transform (DFT) on $A_{MP}(k)$ [19].

Note that although the exact number of reflected paths is unknown, the number of dominating ones is usually no more than four even in an indoor environment [20]. Therefore, it is suggested that the number $L$ in equation (13) is set as 4.

*3) Estimation of $\alpha_i$ and $\theta_i$*

From equations (13) and (15), we arrive at

$$[\mathbf{c}_1 \quad \mathbf{s}_1 \quad \ldots \quad \mathbf{c}_L \quad \mathbf{s}_L]\mathbf{x} = \mathbf{b}$$
(16)

where
$$\mathbf{c}_i = \begin{bmatrix} \cos(2\pi\hat{\tau}_i f(0)) & \ldots & \cos(2\pi\hat{\tau}_i f(K-1)) \end{bmatrix}^T,$$
$$\mathbf{s}_i = \begin{bmatrix} -\sin(2\pi\hat{\tau}_i f(0)) & \ldots & -\sin(2\pi\hat{\tau}_i f(K-1)) \end{bmatrix}^T,$$
$$\mathbf{x} = [x_1 \quad x_2 \quad \ldots \quad x_{2L-1} \quad x_{2L}]^T \text{ with } x_{2i-1} = \alpha_i \cos(\theta_i) \text{ and } x_{2i} = \alpha_i \sin(\theta_i), \text{ and}\ldots$$
$$\mathbf{b} = \begin{bmatrix} A_{MP}(0)/\hat{\bar{A}}(0) & \ldots & A_{MP}(K-1)/\hat{\bar{A}}(0) \end{bmatrix}^T.$$
(17)

Hence, $\mathbf{x}$ can be obtained via least squares. Then we have $\hat{\alpha}_i = \sqrt{x_{2i-1}^2 + x_{2i}^2}$ and $\hat{\theta}_i = \tan^{-1}(x_{2i}/x_{2i-1})$.

### C. Example of Phase Error Correction

Next, we give an example of phase error correction with the proposed scheme in Fig. 2. We assume $f_X(0) = 2400$ MHz, $\Delta f = 1$ MHz, $K = 100$, and one reflected path between the transmitter and receiver with $\tau_{XY} = 20$ ns, $\alpha_{XY} = 0.3$, and $\theta_{XY} = \pi/4$. It is hence observed that the phase error induced by multipath is reduced significantly.

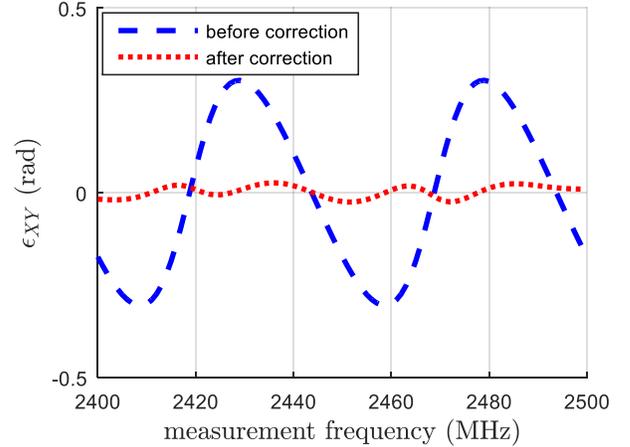

Fig. 2. Phase errors before and after correction

### IV. SIMULATION RESULTS

In this section, we evaluate the performance of our scheme for phase error correction via Monte Carlo simulation experiments. We first investigate the influence of multipath parameters on phase errors for one channel. We then examine the ranging accuracy of RIPS with our scheme for phase error correction.

In simulations, we introduce random noise by defining the phasor of the received signal at the $k$th measurement frequency as $\gamma(k) = A_{XY}(k)\exp(j\varphi_{XY}(k)) + n(k)$, where $n(k)$ is independent and identically distributed zero-mean complex Gaussian noise with a variance of $\sigma^2$ and $k = 0,1,\ldots,K-1$. The signal-to-noise ratio (SNR) is defined as $1/\sigma^2$ and set to 30 dB in the simulation.

System parameters are set as follows: $f_B(0) = 2400$ MHz, $f_A(k) - f_B(k) = 20$ kHz, $\Delta f = 1$ MHz, and $K = 100$.

In each evaluation, 10,000 sets of Monte Carlo experiments are carried out. For comparison, all evaluations are set forth per three separate scenarios: multipath free (MP-free), multipath distorted (MP-distorted), and multipath corrected (MP-corrected). In the MP-free scenario, only random noise is applied; in the MP-distorted and MP-corrected scenarios, multipath is applied along with random noise.

### A. Influence of Multipath Parameters on Phase Errors

Assume that there is one reflected path between each transmitter $X$ and receiver $Y$ with $\tau_{XY} \sim U(5\text{ ns}, 50\text{ ns})$, $\alpha_{XY} \sim U(0.0, 1.0)$ and $\theta_{XY} \sim U(0, 2\pi)$.

Fig. 3 illustrates the RMSE of $\varphi_{XY}$ versus MDR $\alpha_{XY}$. It is apparent from these results that the phase correction process is very effective. The phase error is almost eliminated to the level of random noise when $\alpha_{XY} < 0.4$ (i.e., moderate multipath). Although the phase error increases as $\alpha_{XY}$ grows (multipath becomes more evident), it is remarkable that the MP-corrected scenario outperforms the MP-distorted scenario even under severe multipath conditions (when $\alpha_{XY}$ approaches 1).

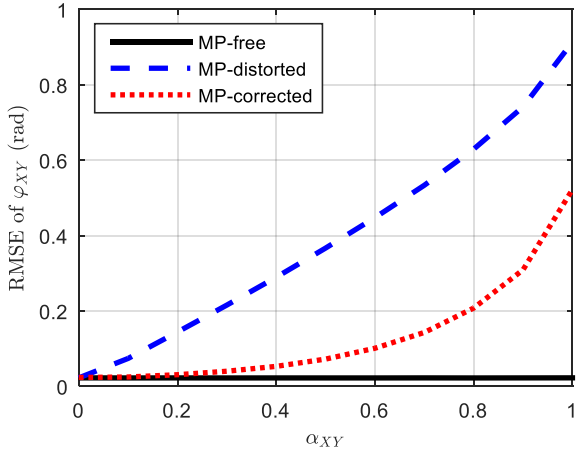

Fig. 3. RMSE of φ$_{XY}$ vs. α$_{XY}$

In Fig. 4, the impact of delay difference $\tau_{XY}$ on error correction is exhibited under moderate multipath with $\alpha_{XY} \sim U(0.1, 0.4)$. The RMSE of $\varphi_{XY}$ for MP-corrected RIPS decreases as $\tau_{XY}$ increases. It is nearly at the level of MP-free RIPS when $\tau_{XY}$ is greater than 10 ns.

To fully understand this phenomenon, a review of multiple tone parameter estimation with DFT is helpful. According to the theory of DFT, the minimum spacing between two resolvable frequencies (i.e., the frequency resolution) in a multiple-tone signal is $1/T = f_s/N$, where $T$ is the observation duration, $f_s$ is the sampling rate, and $N$ is the number of samples. In the estimation of $\tau_{XY}$, $1/\Delta f$ and $K$ act as "sampling rate" and "number of samples," respectively. Moreover, "frequency resolution," $1/(K\Delta f)$, is actually the "delay resolution" (i.e., the difference in delays between the reflected path and the direct one). Note that $K\Delta f$ is the bandwidth of the measurement frequencies; a wider bandwidth of measurement frequencies translates to a finer delay resolution. Given that $\Delta f = 1$ MHz and $K = 100$, the delay resolution is 10 ns. This explains the excellent performance of the MP-corrected scenario in Fig. 4 when $\tau_{XY}$ is more than 10 ns.

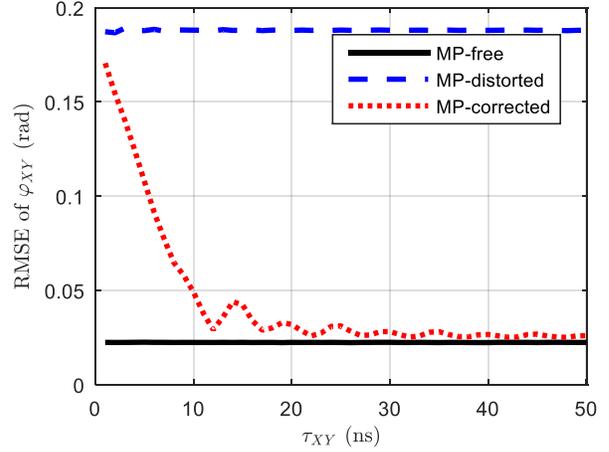

Fig. 4. RMSE of φ$_{XY}$ vs. τ$_{XY}$

### B. Ranging Performance of RIPS with Phase Error Correction

Finally, we show the effect of the phase correction method on distance (q-range) estimation. Assume that $d_q = 75$ m, $\alpha_{XY} \sim U(0.0, 1.0)$, and $\tau_{XY} \sim U(10\text{ ns}, 50\text{ ns})$. Fig. 5 shows the corresponding cumulative density function (CDF) of the absolute errors in the q-range estimation. For MP-distorted RIPS, the median and 95[th] percentile errors are 0.33 m and 7.4 m, respectively. With phase correction, they drop to 0.05 m and 0.38 m, respectively. These results confirm the effectiveness of the multipath error correction method for RIPS.

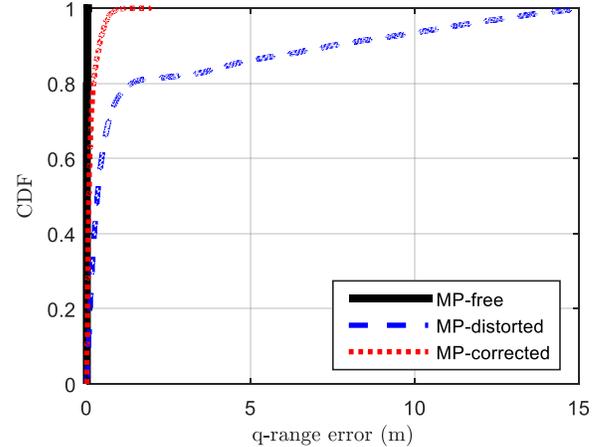

Fig. 5. CDF of q-range error

## V. CONCLUSIONS

In this letter, we propose to utilize the untapped amplitude information in ranging signals of RIPS for multipath mitigation. Through this study's proposed scheme for multipath error correction, it was shown that errors in the realms of phase measurement and range estimation can be significantly reduced under both moderate and severe multipath conditions.

Moreover, the proposed approach entails the following additional advantages: 1) it avoids inconvenient system parameter design choices, and 2) it does not compromise any support for multiple carrier frequencies.